\documentclass[twocolumn,groupedaddress,amsmath,aps,longbibliography,nofootinbib]{revtex4-2}
\usepackage{graphicx}% Include figure files
\usepackage{hyperref}
\usepackage{bm}% bold math
\usepackage{xcolor}
\usepackage{physics}

\begin{document}

\title{Thermoelectric performance of Ni-Au metallic alloys determined by resonant scattering}
% Force line breaks with \\

\author{Kacper Pryga}
\affiliation{AGH University of Krakow, Faculty of Physics and Applied Computer Science, Aleja
Mickiewicza 30, 30-059 Krakow, Poland}
\author{Bartlomiej Wiendlocha}
\email{wiendlocha@fis.agh.edu.pl}
\affiliation{AGH University of Krakow, Faculty of Physics and Applied Computer Science, Aleja
Mickiewicza 30, 30-059 Krakow, Poland}

\date{\today}

\begin{abstract}
This work presents a theoretical study of the electronic structure and transport properties of Ni-Au alloys, recently identified as excellent thermoelectric metals with a power factor significantly exceeding that of conventional semiconductor thermoelectrics.
Using first-principles calculations based on the Korringa-Kohn-Rostoker method combined with the coherent-potential approximation (KKR-CPA) and the Kubo-Greenwood formalism, we demonstrate the key role of resonant scattering in determining the thermoelectric properties of these alloys.
This is supported by calculated densities of states, Bloch spectral functions, electrical conductivity, and thermopower.
Alloying Ni with Au not only induces resonant scattering but also leads to the formation of a flat band below the Fermi level. The combination of these two features results in high thermopower, arising from a transition from resonant to weak scattering regimes near the Fermi level.
Our findings are further compared with analogous calculations for constantan, a Ni-Cu alloy long regarded as a reference thermoelectric metal.
We show the key differences between the Ni-Au and Ni-Cu systems that explain why Ni-Au exhibits nearly twice the thermopower of Ni-Cu.
Finally, we simulate the effect of lattice parameter variation on the thermoelectric performance of Ni-Au and suggest that this is a promising pathway for further enhancement, for example through additional alloying or layer deposition.

% \begin{description}
% Secondary publications and information retrieval purposes.
% You may use the \texttt{description} environment to structure your abstract;
% use the optional argument of the \verb+\item+ command to give the category of each item. 
% \end{description}
\end{abstract}


\maketitle

% ################################################\
\section{Introduction}
% ################################################\

In the field of thermoelectric (TE) materials, scientific interest has focused for years on semiconductors, while metallic alloys were often, apart from a few examples, omitted from the spotlight.
The efficiency of TE materials is defined using the dimensionless figure of merit $ZT=\frac{S^2\sigma}{\kappa}T$, where $S$, $\sigma$, $\kappa$, $T$, are thermopower, electrical conductivity, thermal conductivity, and temperature. In most cases, the Seebeck coefficient of metals is at least an order of magnitude lower than in semiconductors, and considering that conversion efficiency depends on $S^2$, the apparent advantage of semiconductors is justified \cite{Mahan1996,Biswas2022,Goldsmid2010IntroductionThermoelectricity,Li2019SnTe-BasedThermoelectrics,Zevalkink2018,Lee2022}. 
Furthermore, in semiconductors, the reduction of the lattice thermal conductivity $\kappa_l$ has been extensively used for improvement of $ZT$ (e.g., through phonon scattering on point defects or nanoparticles \cite{nanoparticles,Lyu2024}, grain boundaries  \cite{Cahill2014,grain-calc}, introducing dislocations \cite{Hanus2021} or nanostructuring \cite{nanograins,Zheng2021,thermal-si-alloys,Stamper2024}), while in metallic compounds, a large electronic contribution to thermal conductivity impedes this approach, resulting in lower thermoelectric efficiency and decreased interest from the perspective of TE power generation.
However, for active cooling applications, materials with a high power factor and thermal conductivity are necessary~\cite{zebarjadi2015,adams2019,zebarjadi2022}, making metallic alloys possibly a better choice.

Among the metallic alloys with good thermoelectric performance, one can mention the nickel-copper alloy (constantan). 
It has a thermopower of $S=-45$ $\mu$V/K at 300 K for Ni$_{0.4}$Cu$_{0.6}$ \cite{Ho1978} and has been used for decades in thermocouples. 
Similar TE properties are found among other transition metal-noble metal alloys, including Pd-Ag or Pd-Au \cite{Ho1993}. Other metallic systems with noticeable thermopower are found among pure metals such as Co ($\abs{S}\sim30$ $\mu$V/K ) \cite{Vedernikov1969,Watzman2016} or Yb \cite{Vedernikov1969}, heavy-fermion compounds such as YbAl$_3$ \cite{Rowe2002}, YbAl$_3$:Sn \cite{Li2014} or CePd$_3$-based compounds \cite{Boona2012} that possess one of the best low temperature thermoelectric efficiency among metallic compounds, with $ZT$ reaching 0.3 at room temperature. 
However, usage of alloys such as Pd-Ag or Pd-Au is limited because of the high price, despite the fact that they exhibit relatively good thermoelectric properties. In addition, to spark widespread application of metallic alloys in TE applications, their efficiency has to be improved.

In a recent study \cite{Garmroudi2023} Ni$_x$Au$_{1-x}$ alloy system was discovered to possess exceptional thermoelectric capabilities. The best sample at 43\% Ni content achieved the power factor $PF=\sigma S^2$ up to $34$ mWm$^{-1}$K$^{-2}$, vastly outperforming any bulk metallic or semiconducting material. Similarly, the figure of merit reached $ZT\sim 0.45$ at room temperature, among the highest for metallic systems.

In this work, we present a theoretical analysis of both the electronic structure and transport properties of Ni$_x$Au$_{1-x}$ metallic alloy using \textit{ab-initio} calculations and explain the reason for its remarkable thermoelectric capabilities.
Our treatment goes beyond the phenomenological $s-d$ scattering model used in \cite{Garmroudi2023} and shows that resonant scattering below the Fermi level in Ni$_x$Au$_{1-x}$, in combination with a sharp linear band above $E_F$, determines its extraordinary performance. This analysis is presented as a side-by-side study with Ni-Cu alloy, and it explains why Ni-Au became a better thermoelectric material than constantan, despite the even better-looking density of states profile of the latter. 
The difference comes mainly from the {\it weaker} scattering of conduction electrons with energies at and above the Fermi level in Ni-Au than in Ni-Cu, highlighting the importance of scattering in parallel with the density of state modifications due to alloying.
Identifying this cause shows the directions to look for better metallic thermoelectric alloys.

% ################################################\
\section{COMPUTATIONAL METHODS}
% ################################################\
% \printinunitsof{in}\prntlen{\textwidth}

Electronic structure calculations were performed using the full-potential Korringa-Kohn-Rostoker (KKR) method with coherent potential approximation (CPA) as implemented in the Munich {\sc SPR-KKR} package \cite{Ebert2011CalculatingApplications,Ebert2019TheMunich}.
The Vosko, Wilk, and Nussair parametrization \cite{Vosko1980AccurateAnalysis} of the local density approximation (LDA) was used for the crystal potential and the Lloyd formula was used to establish the Fermi level position \cite{Ebert2011CalculatingApplications}. 
The spin-orbit interaction was included in the calculations, and the angular momentum cutoff was set to $l_{\rm max}=3$. A dense mesh of $\sim5.7\cdot 10^3$ $\vb{k}$-points within the irreducible part of the Brillouin zone was used for the self-consistent-field cycle and $\sim3\cdot 10^5$ $\vb{k}$-points for the density of states (DOS) and Bloch spectral density functions (BSFs) calculations. For the computation of the energy-dependent electric conductivity $\sigma(E)$ (called here the transport function), a mesh of $\sim1.5\cdot 10^{4}-1.2\cdot10^6$ $\vb{k}$ points was used, with convergence tests performed on meshes at least 1.5 times denser (examples of convergence tests are shown in Supplemental Material~\cite{suppl}). 
The energy range of $-0.86 < E-E_F < 0.86$ eV was considered, with an energy step of 0.1 mRy. The experimental crystal structure \cite{Garmroudi2023} was used in the calculations, in addition to variants with the 43\% and 80\% Ni content, in which the lattice parameters were adjusted according to Vegard's law.

To describe the electronic band structure of the alloys studied, the Bloch spectral density functions were calculated, which are a generalization of the electronic dispersion relations for the disordered system \cite{Ebert2011CalculatingApplications,Ebert1997RelativisticAlloys,Faulkner1980CalculatingApproximation}. 
In the case of an ordered material, BSFs take the form of a Dirac delta: $A^B(\vb{k},E)=\delta(E-E_{\nu,\vb{k}})$ describing the position of the electronic bands $E_{\nu,\vb{k}}$ (with $\nu$ being the band index, $\vb{k}$ the wave vector and $E_{\nu,\vb{k}}$ the energy eigenvalue).
In the disordered system, where alloy scattering occurs, bands become smeared. In many cases, BSF at given $\vb{k}$-point can be then described by a Lorentz function~\cite{Gyorffy1979FirstAlloys,Gordon1981OnAlloys,PhysRevB.29.4217,Butler1985}:
\begin{equation}
    A^B(\vb{k},E)=L(E)=\dfrac{1}{\pi} \dfrac{\frac{1}{2}\Delta}{\left(E-E_0\right)^2+\left(\frac{1}{2}\Delta\right)^2},
    \label{eq:lorentz}
\end{equation}
where the location of the peak indicates the band center and its full width at half maximum $\Delta$ is associated with the lifetime of the electronic state:
\begin{equation}
    \tau=\hbar/\Delta.
\end{equation}
Generally, the stronger the alloy scattering, the wider the BSF and the larger $\Delta$. 
For a very strong and resonant scattering~\cite{Gyorffy1979FirstAlloys,Wiendlocha2013FermiCalculations,Wiendlocha2021ResidualSemiconductors}, the BSF may lose the Lorentzian shape, making $\tau$ a not well-defined quantity.

In studies of thermoelectric properties, two approaches were used.
First, in combination with the KKR-CPA method and the Kubo-Greenwood formalism \cite{Kubo1957,Greenwood1958,Butler1985,Swihart1986}, the energy-dependent electrical conductivity $\sigma(E)$ (the transport function) of the ground state ($T = 0$ K) was calculated in SPR-KKR \cite{Ebert2019TheMunich,Ebert2011CalculatingApplications,Kdderitzsch2011,Popescu2017}. 
This formalism allows us to study the electron scattering effects induced by the disordered set of electronic potentials and to include the vertex corrections. 
Because of the strong and resonant nature of this scattering in Ni-Au, 
it becomes the dominant scattering mechanism, especially for larger Ni concentrations. Furthermore, because of the rapid energy dependence of $\sigma(E)$ set by resonant scattering, it also gives the dominant contribution to thermopower $S$. 
By neglecting phonon effects on $\sigma(E)$, thermopower is directly calculated without any adjustable parameters based on the standard approach \cite{Blatt1976}:
\begin{equation}
    S(T)=-\dfrac{1}{eT} \dfrac{L^{(1)}}{L^{(0)}},\label{eq:seebeck}
\end{equation}
where:
\begin{equation}
    L^{(n)}=\int_{E_{\rm min}}^{E_{\rm max}} dE \qty(-\pdv{f}{E})(E-\mu)^n\sigma(E).\label{eq:ln}
\end{equation}
Here, $e$, $f$, and $\mu$ denote electronic charge, Fermi-Dirac distribution function, and chemical potential, the latter being calculated for a given $E_F$ position and temperature. 
The definite integral is taken in the symmetrical energy range around the Fermi energy, in our case $E_{\rm min} = E_F - 0.86$ eV,  $E_{\rm max} = E_F + 0.86$ eV.
A similar approach has been successfully used in studies of thermopower and transport properties in various thermoelectric materials \cite{Banhart1995,Wiendlocha2018Thermopower}, including metallic alloys such as Ag-Pd, Au-Pd \cite{Banhart1995,PhysRevB.29.4217} and Ni-Cu \cite{Wiendlocha2018Thermopower,Vernes2003}, and the common feature of these systems was the presence of resonant energy states of impurity atoms, which introduce the dominant scattering mechanism.
Other scattering processes that may contribute to thermopower, such as electron-(para)magnon scattering, as well as all electron-phonon-related processes ~\cite{nielsen74,vilenkin79,vilenkin80,gallagher85,kaiser84, mahan90,livanov93}(including electron-phonon renormalization of mass, relaxation times, and velocity, which lead to enhancement of thermopower, and electron-phonon-impurity interference contribution that partially cancel this enhancement~\cite{livanov93}) are neglected. 
As discussed later, for lower Ni concentrations, our results show an underestimate of $S$ at lower temperatures, which may be related to neglected electron-phonon enhancement of thermopower, and an overestimation of $S$ at high temperatures, probably due to neglected phonon influence on $\sigma(E)$, but overall accuracy is satisfactory.

To show how resonant scattering is important here, we also performed calculations using Boltzmann transport theory in the constant relaxation time approximation (CRTA), as implemented in the {\sc BoltzTrap} code \cite{boltztrap}. The electronic structure used in the CRTA calculation was obtained from supercell calculations using the all-electron full potential augmented plane wave method plus local orbitals (APW+lo), within the package {\sc WIEN2k} \cite{wien2k,Blaha2020WIEN2k:Solids}. These calculations also included the spin-orbit interaction.
To simulate the alloy structure, a 3$\times$3$\times$3 supercell was constructed, $\mathrm{Ni_{11}Au_{16}}$, corresponding to $\mathrm{Ni_{0.407}Au_{0.593}}$ and using the experimental lattice parameter \cite{Garmroudi2023}.
In case of self-consistent cycle and eigenvalue calculations, a dense grid of nearly $\sim1.9\cdot 10^{5}$ $\vb{k}$-points per primitive cell was used, with a total energy convergence of $10^{-4}$~Ry. The muffin tin radii were set at 1.322~\AA\ with valence and core separation of $–6$ Ry.   

\begin{figure}[t]
    \includegraphics[width=0.95\linewidth]{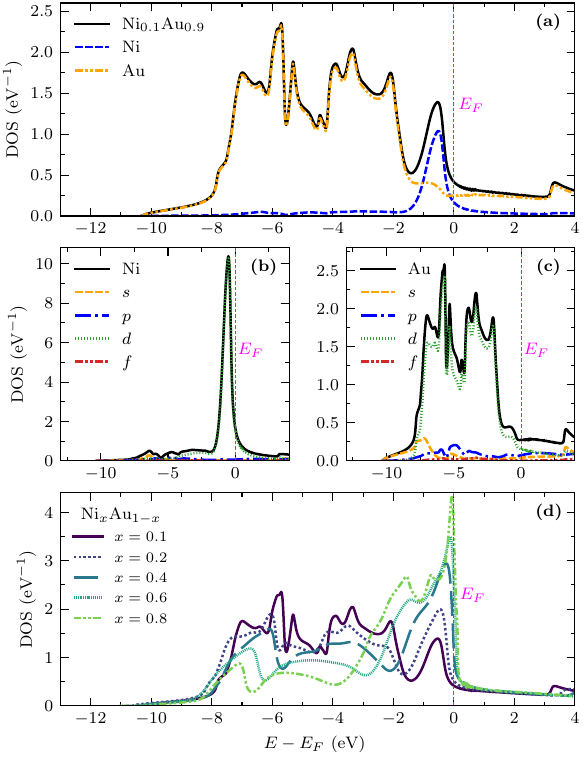}
    \caption{(a) Calculated density of states (DOS) of $\mathrm{Ni_{0.1}Au_{0.9}}$ with partial density of states in panel (b) and (c); (d) evolution of total DOS of Ni$_x$Au$_{1-x}$ with increasing Ni content.}
    \label{fig:ni10au90-dos-pdos}
\end{figure}

\begin{figure*}[htb!]
    \includegraphics[width=0.95\textwidth]{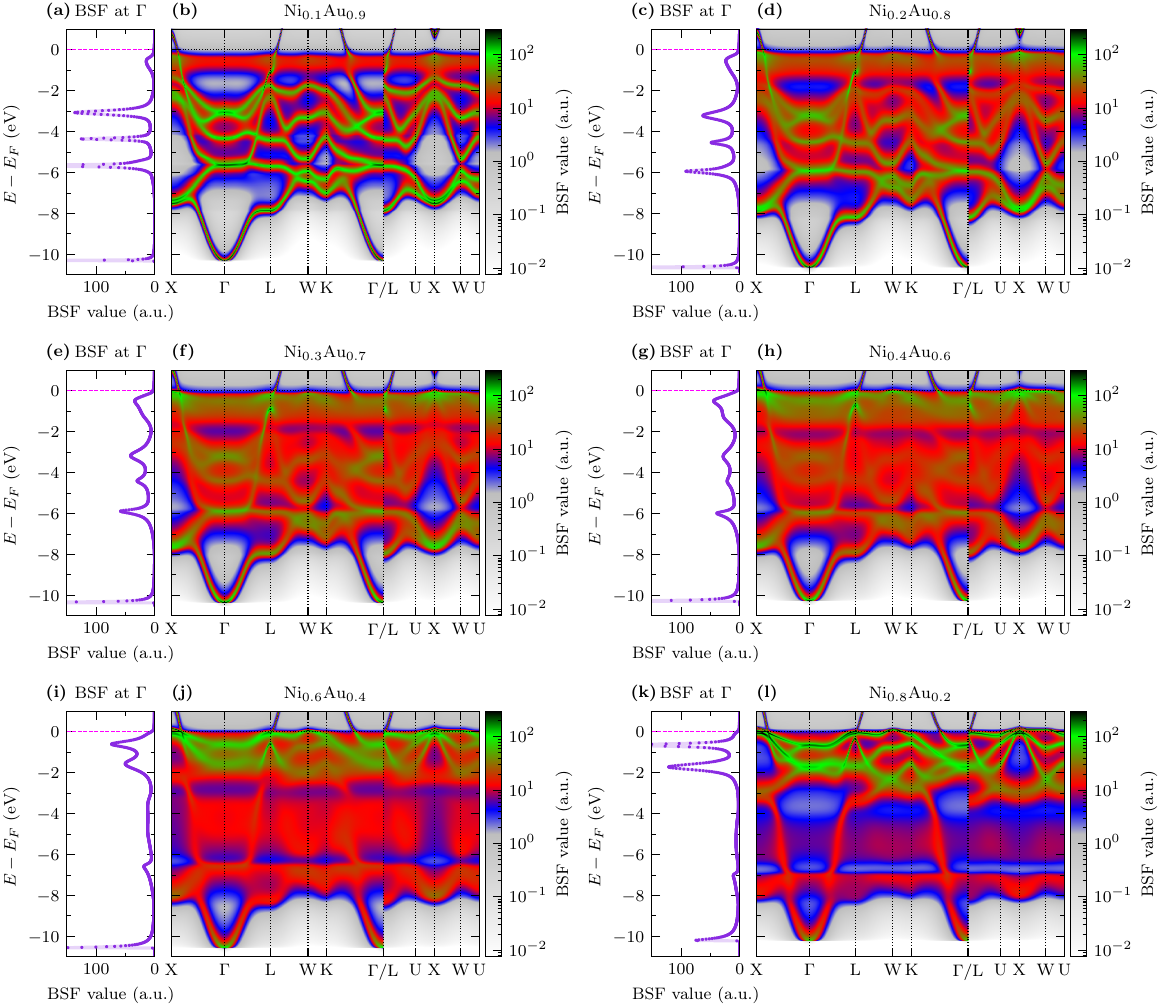}
    \caption{Plots of Bloch spectral functions at $\Gamma$ point (with light violet line as a guide to the eye) as well as 2D projections of BSFs for (a,b) $\mathrm{Ni_{0.1}Au_{0.9}}$, (c,d) $\mathrm{Ni_{0.2}Au_{0.8}}$, (e,f) $\mathrm{Ni_{0.3}Au_{0.7}}$, (g,h) $\mathrm{Ni_{0.4}Au_{0.6}}$, (i,j) $\mathrm{Ni_{0.6}Au_{0.4}}$ and (k,l) $\mathrm{Ni_{0.8}Au_{0.2}}$. In case of 2D projections (panels b,d,f,h,j,l) color represents BSF value in logarithmic scale with black color corresponding to values over 300 atomic units.     
    }
    \label{fig:bsf_overview}
\end{figure*}

% ################################################\
\section{Results and discussion}
\subsection{Electronic structure}
% ################################################\

\begin{figure}[t]
    \includegraphics[width=1\linewidth]{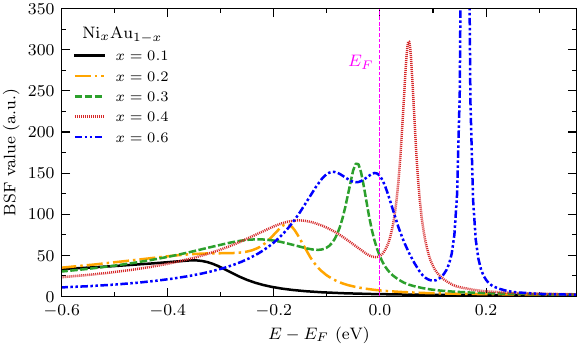}
    \caption{Calculated BSFs of Ni$_x$Au$_{1-x}$ in the vicinity of the Fermi level for the X high symmetry point for several Ni concentrations. We see a gradual formation of virtual band as sharpening of the spectral function. Above 40\% a two-peak structure is observed as two bands at X can be distinguished in Fig.~\ref{fig:bsf_overview} in this energy range.}
    \label{fig:bsf_x_rebuild}
\end{figure}

Fig.\,\ref{fig:ni10au90-dos-pdos}(a-c) shows total and partial density of states for Ni$_{0.1}$Au$_{0.9}$. At this concentration, the total DOS consists mainly of the Au 5$d$ states; however, a large DOS peak of Ni 3$d$ states appears just below the Fermi level. This indicates a possibility of the formation of a resonant level (RL), just as in the Ni-Cu constantan case~\cite{Gyorffy1979FirstAlloys,Gordon1981OnAlloys,Wiendlocha2018Thermopower}.
Although the Ni-Au alloy is ferromagnetic above $\sim55\%$ of the Ni content \cite{Kuentzler1979}, 
samples up to 70\% possess Curie temperatures below $\sim 300$ K \cite{Bergmann1976,Bergmann1977,Kuentzler1979}, which is lower than our area of interest from the perspective of thermoelectric properties. Because of that, all electronic structure calculations presented in this section are done for a non-magnetic state, even for $x > 0.5$. 
Spin-polarized DOS for Ni$_{0.5}$Au$_{0.5}$ is additionally shown in Supplemental Material~\cite{suppl}. 

Further alloying with nickel affects mainly the resonant peak, which becomes larger, steeper, and more asymmetric, as shown in Fig.\,\ref{fig:ni10au90-dos-pdos}(d). 
This behavior was also observed in Ref. \cite{Garmroudi2023}, where ab-initio results were shown, however at lower energies (below $-2$ eV) our results visibly differ from those of Ref. \cite{Garmroudi2023}.  This might stem from different computational details, such as different parametrizations of the LDA exchange-correlation potential or different treatments of relativistic effects.
Other DFT studies of the Ni-Au system~\cite{Hao2021,Karmakar2022} were focused on surfaces, so direct comparison is not available. 
In the remaining energy range, DOS is generally unaffected by Ni alloying, particularly above 1 eV above $E_F$, where the DOS is almost identical in all the concentrations studied.

\begin{figure*}[htb!]
    \includegraphics[width=1\textwidth]{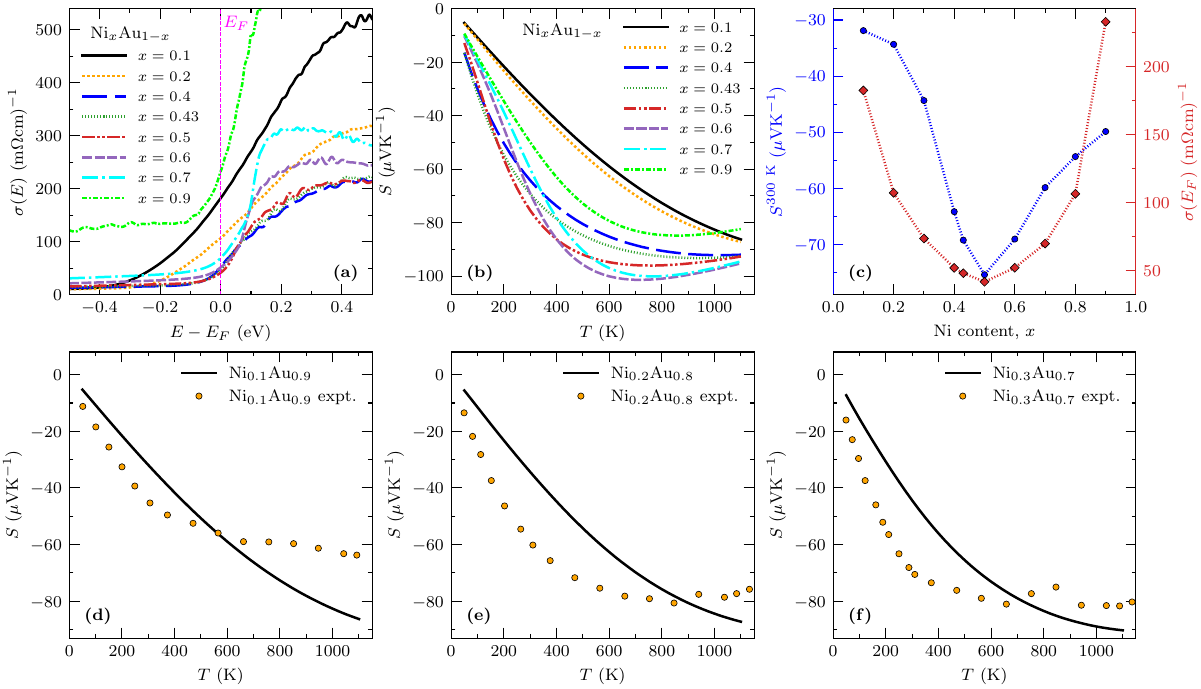}
    \caption{Transport properties of Ni$_x$Au$_{1-x}$ for $x=0.1-0.9$. (a) Energy-dependent conductivity $\sigma(E)$; (b) calculated Seebeck coefficient as a function of temperature $S(T)$; (c) thermopower at 300 K $S^{300\ \mathrm{K}}$ and conductivity at Fermi level $\sigma(E_F)$ as a function of Ni content. Panels  (d-f) show comparison between experimental~\cite{Garmroudi2023} and calculated temperature-dependent Seebeck coefficient for chosen concentrations. In case of the Ni$_{0.3}$Au$_{0.7}$ shown in panel (f), experimental points obtained during heating are shown, as due to phase segregation occurring at high temperatures results upon cooling did not correspond to single-phase material~\cite{Garmroudi2023}.}
    \label{fig:transport_overview}
\end{figure*}

Electronic dispersion relations are presented in the form of two-dimensional projections of the Bloch spectral density functions in Fig.\,\ref{fig:bsf_overview}, with the color corresponding to the value of BSF. In addition, the BSFs are plotted for the $\Gamma$ $\vb{k}$-point to further highlight the trend of how the smearing of the bands and the BSF shape change with alloying in different energy ranges.

Starting with Ni$_{0.1}$Au$_{0.9}$ in Fig.\,\ref{fig:bsf_overview}(a,b), the band structure is clearly divided into three qualitatively different regions. 
The first, with the lowest energies between $-10$ eV and $-2$ eV, has relatively sharp bands, originating from the smeared band structure of Au. In Fig.\,\ref{fig:bsf_overview}(a), at $\Gamma$ point, they appear as narrow Lorentzian-like peaks. 
In the second region between -2 eV and the Fermi level, two important modifications occur. First, a new feature appears in the band structure, not present in pure Au~\cite{band-gold}: 
a flat and strongly smeared virtual band, that continues through a full path of the Brillouin zone included in Fig.\,\ref{fig:bsf_overview}. At the $\Gamma$ point in panel (a), it appears as a broad hump in the BSF, centered around $-0.5$ eV. This is formed by the resonant state associated with Ni, responsible for the peak in the partial Ni DOS, seen in Fig.\,\ref{fig:ni10au90-dos-pdos}(a).
Second, this resonant state distorts and smears the sharp and almost linear band of Au, which continues, e.g. between K and $\Gamma$ or from the L-point at about $-2$ eV up to the Fermi level.
This linear band continues further with increasing energy, crosses $E_F$ and goes above, defining the third characteristic region, from 0 to 1 eV in Fig.\,\ref{fig:bsf_overview}. In this region, which is outside the resonance, band smearing is much weaker.
These three band-structure regions evolve with increasing Ni concentration, and this evolution is responsible for the thermoelectric performance of the Ni-Au alloy. 

At 20\% Ni, further smearing of the lowest bands is observed, which is expected due to the increasing level of alloying, as electron scattering becomes more prominent. 
This corresponds to the broadening of the BSFs (Fig.\,\ref{fig:bsf_overview}(c)) in the lowest energy region. 
In parallel, in the second energy region ($-2$ eV to $E_F$) the intensity of the resonance virtual band begins to grow, which is indicated, e.g. by the appearance of bright green areas near the X $\vb{k}$-point. Furthermore, the linear band at L or between K and $\Gamma$, due to hybridization with the resonance, begins to split into two parts.
In the third energy region (above $E_F$), only a small increase in band smearing can be observed, which will be evidenced in the discussion of conductivity, as it is not well seen on the scale of this figure.

The above-mentioned trends continue when $x$ increases. 
For Ni$_{0.3}$Au$_{0.7}$ (Fig.\,\ref{fig:bsf_overview}(e,f)) and Ni$_{0.4}$Au$_{0.6}$ (Fig.\,\ref{fig:bsf_overview}(g,h)) in the first region deep below $E_F$, the bands are now very strongly distorted, which is also indicated by the highly non-lorentzian BSFs shape in Fig.\,\ref{fig:bsf_overview}(e,g). 
In region 2 just below $E_F$, the resonance virtual band continues to form and even slightly sharpen, but remains very strongly smeared. It is especially well seen at the X point, where it starts slightly above $E_F$, and continues in the $\mathrm{X}-\Gamma-\mathrm{L}-\mathrm{W}$ path with a lowest energy center at about $-0.5$ eV at the $\Gamma$ point. The additional virtual band starts to be visible for the Ni$_{0.4}$Au$_{0.6}$ just below the first one.
In parallel, due to further hybridization of the Ni resonance with the Au states, the linear band disconnects around $-0.5$ eV. Its upper part joins the upper virtual band, which is well seen e.g. at L and in the $\mathrm{K}-\Gamma$.
In the third region, no qualitative changes are visible.

The virtual band formation is additionally shown in Fig.\,\ref{fig:bsf_x_rebuild}, where BSFs at the X $\vb{k}$-point are plotted for several Ni concentrations. 
Here, between 20\% and 40\% of Ni a new peak arises from the hump at $-0.4$ eV originating from the resonant Ni level. 
For higher Ni concentrations (where, however, the single-phase material was not obtained in ~\cite{Garmroudi2023}) 
the virtual bands in region 2 sharpen further (Fig.\,\ref{fig:bsf_overview}(i-l)). The band structure becomes more Ni-like, with more dispersive and well-defined bands seen in Fig.\,\ref{fig:bsf_overview}(k-l). The original, Au-dominated part of the band structure in region 3 becomes less pronounced and more smeared. 

% ################################################
\subsection{Transport properties}
% ################################################

\begin{figure}
    \includegraphics[width=1\linewidth]{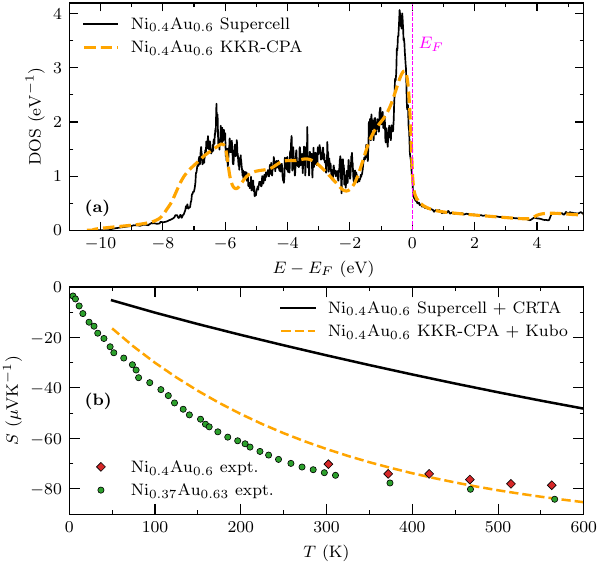}
    \caption{(a) Comparison of DOS of Ni$_{11}$Au$_{16}$ supercell ($\mathrm{Ni_{0.407}Au_{0.593}}$) (APW+lo method) with the KKR-CPA. (b) Comparison of calculated temperature-dependent Seebeck coefficients from the supercell method with the constant relaxation time approximation (CRTA), which neglect electron scattering, with KKR-CPA and Kubo formalism, where disorder-induced electron scattering is taken into account. CRTA strongly underestimates the absolute value of thermopower. In addition, experimental results for Ni$_{0.37}$Au$_{0.63}$ and Ni$_{0.4}$Au$_{0.6}$ from Ref.\,\cite{Garmroudi2023} are shown and agree rather well with the  KKR-CPA + Kubo calculations confirming that the resonant scattering is the dominating feature leading to a high thermopower. Experimental data are shown for measurement obtained only during heating process due to phase segregation occurring at high temperatures.
    }
    \label{fig:dos_transport_ni0.4au0.6_csta}
\end{figure}

% Transport properties calculations were performed using the combination of the KKR-CPA and Kubo-Greenwood formalism. In this approach, it is possible to calculate the energy-dependent electrical conductivity of the ground state ($T = 0$ K) $\sigma(E)$ (the transport function) without any adjustable parameters, since electron scattering due to the presence of randomly distributed impurity atoms is taken into account. 
% The thermopower $S$ is then calculated based on the standard approach \cite{Blatt1976}:
% \begin{equation}
%     S=-\dfrac{1}{eT} \dfrac{L^{(1)}}{L^{(0)}},\label{eq:seebeck}
% \end{equation}
% where:
% \begin{equation}
%     L^{(n)}=\int dE \qty(-\pdv{f}{E})(E-\mu)^n\sigma(E).\label{eq:ln}
% \end{equation}
% Here, $e$, $f$, and $\mu$ denote electronic charge, Fermi-Dirac distribution function, and chemical potential, the latter being calculated at each given $E_F$ position and temperature.
% As electron-phonon scattering is neglected here, this methodology can predict the correct thermopower values in materials where electrons are strongly scattered by the disordered set of atomic potentials, that is, especially in the case of alloys where resonant scattering occurs.
% This approach has been successfully used in studies of transport properties and thermoelectric calculations in various thermoelectric materials \cite{Wiendlocha2018Thermopower,Wojciechowski2020,Banhart1995}, including metallic alloys such as Ag-Pd, Au-Pd \cite{PhysRevB.29.4217,Banhart1995} and Ni-Cu \cite{Wiendlocha2018Thermopower,Vernes2003}. 

Figure\,\ref{fig:transport_overview}(a) shows the calculated transport function $\sigma(E)$ as a function of Ni concentration. 
In general, we observe small conductivity values below $E_F$, which then increase when the energy crosses the Fermi level. 
Rapidly changing conductivity is certainly beneficial to obtain high thermopower, as predicted by the Mott formula $S \propto -\dd\ln \sigma(E)/\dd E$.
Let us now analyze how, in the context of the evolution of the BSFs, described above, this rapidly changing $\sigma(E)$ function of Ni$_{1-x}$Au$_x$ is understood and how it changes with $x$. 
First, for 10\% Ni, the appearance of resonance leads to smearing of the linear band below $E_F$ (energy region 2 in the BSF discussion), resulting in a decrease in $\sigma(E)$ of the linear band below $E_F$, but keeping its high conductivity above $E_F$ (region 3).
As the concentration of Ni increases from 10\% to 50\%, as observed in the BSFs, in region 2 the flat and strongly smeared virtual band just below $E_F$ is formed and the linear band near $-0.5$ eV is broken. Its upper- and lower-lying parts are being merged into flatter, smeared, and less-conducting virtual bands. 
This explains the gradual decrease of $\sigma(E)$ below $E_F$, which results in an S-shaped profile.
Above the Fermi level, a slight increase in band smearing results in a slowly reduced transport function until $x$ reaches 30\%. Then this effect saturates, since for compositions between 30\% and 50\% Ni, the alloy has similar $\sigma(E)$ values (the graphs of $\sigma(E)$ for all concentrations are shown in the Supplemental Material~\cite{suppl}). 
In addition, the sharpening of the virtual band with increasing $x$ explains the small increase in $\sigma(E)$ below $-0.4$ eV between 10\% and 50\% (see also a zoom of Fig.\,\ref{fig:transport_overview}(a) in  Supplemental Material~\cite{suppl}).

When the Ni content exceeds $x=0.5$, the transport function starts to increase strongly in the entire covered energy range as
a consequence of further sharpening of the spectral functions, formation of the more dispersive and Ni-like band structure and increase of the total DOS around $E_F$.

This suggests that for lower temperatures, where mainly states very close to $E_F$ contribute to the thermoelectric transport properties, the highest thermopower can be expected at $x = 0.5$. For Ni$_{0.6}$Au$_{0.4}$ $\sigma(E)$ has a much steeper slope above $E_F$ in the $0-0.2$ eV range, however, the values below $E_F$ are $\sim2$ times higher compared to samples below $x=0.5$. 

The thermopower computed as a function of temperature for different concentrations of Ni is presented in Fig.\,\ref{fig:transport_overview}(b), and the value for $T = 300$~K as a function of $x$ in Fig.\,\ref{fig:transport_overview}(c).
Thermopower is negative for all temperatures and compositions, which is directly related to the monotonically increasing $\sigma(E)$ function, resulting in a positive sign of $\dd\ln \sigma(E)/\dd E$. 
It is instructive here to mention the qualitative difference between this kind of metallic alloy and typical semiconductor, where $\sigma(E) \propto N(E)$, with $N(E)$ being the density of states. A semiconductor with a shape of $N(E)$ as in Fig.,\ref{fig:ni10au90-dos-pdos} and $E_F$ located on a decreasing would have a positive thermopower. 
Here, as is characteristic for metals with $s-d$ scattering, $\sigma(E) \propto 1/N_d(E)$ \cite{Banhart1995,Garmroudi2023}, where $N_d$ is the $d$-orbital density od states, due to the prime importance of carrier scattering, with $\tau^{-1} \propto N_d$.
This results in a negative thermopower.

In terms of absolute values, the room temperature thermopower is the highest for $x = 0.5$, where we can expect the strongest disorder-induced resonant scattering, with $S=-75.4$~$\mu$V/K.
Despite saturation of $\sigma(E)$ above $E_F$ with Ni content up to 50\%, a constant increase in $S(T)$ is present for those concentrations in the whole temperature range (Fig.\,\ref{fig:transport_overview}(b)). 
As for Ni$_{0.43}$Au$_{0.57}$ -- the best experimental sample, our results suggest the maximal value of the Seebeck coefficient of $S=-93.2$~$\mu$V/K at 925~K. With further doping with Ni, the thermopower increases at higher temperatures, achieving $-101.4$ $\mu$V/K for Ni$_{0.6}$Au$_{0.4}$ at $\sim725$~K at the cost of deterioration in low temperature performance. 
Additionally, in Fig.\,\ref{fig:transport_overview}(c) the $\sigma(E_F)$ (inverse of a residual resistivity) is shown. As one can notice, the least-conductive composition is the one that exhibits the best room-temperature thermopower.

The comparison of the calculated and experimental thermopowers as a function of temperature is given in Fig.\,\ref{fig:transport_overview}(d-f) for $x = 0.1 - 0.3$ and in Fig.\,\ref{fig:dos_transport_ni0.4au0.6_csta}(b) for $x = 0.4$. 
For $x\geq0.3$ experimental points obtained only during the heating of the samples are shown due to phase segregation at higher temperatures~\cite{Garmroudi2023}.
The general temperature and composition dependence of $S$ is reproduced satisfactorily. The deviations between the calculated and measured values appear most likely due to the neglect of electron-phonon interactions in the calculations of $\sigma(E)$. This correlates with the smallest deviations observed for $x = 0.4$, where the resonant scattering is stronger (i.e. more dominant) than for $x = 0.1 - 0.3$.
In the low to mid-temperature range, in all cases, the calculated Seebeck coefficient is underestimated. This may indicate the presence of an electron-phonon enhancement of the thermopower, which is not taken into account in our calculations. As shown in ~\cite{kaiser84,gallagher85}, this enhancement decreases with increasing temperatures above the Debye temperature, which is of the order of 200 K in Ni-Au alloys~\cite{eno1977specific}.
The electron-phonon coupling parameters $\lambda$, estimated based on the measured electronic specific heat coefficient $\gamma$ ~\cite{eno1977specific} and our calculated values of $N(E_F)$, are between 0.12 and 0.46 (see Supplemental Material~\cite{suppl}), therefore not large but not negligible. 
On the other hand, at high temperatures, the absolute values of the calculated thermopowers are larger than the experimental ones. This can be understood as the large $S$ results from the steep $\sigma(E)$ function, which rapidly changes from the resonant to the weakly scattering energy regions. Electron-phonon scattering, which becomes strong at high temperature, is expected to flatten the $\sigma(E)$ function by reducing conductivity in the high-conductivity energy region, in which the alloy scattering is weaker.

% Also, note that 

To highlight the importance of the energy-dependent resonant electron scattering in determining the thermopower of Ni-Au, we have additionally calculated the thermopower using the constant relaxation-time approximation (CRTA). 
For this approach, the electronic structure of 3$\times$3$\times$3 supercell Ni$_{11}$Au$_{16}$ (which corresponds to Ni$_{0.407}$Au$_{0.593}$) was calculated using the {\sc WIEN2k} package and used as input to the CRTA calculations in the {\sc BoltzTraP} code \cite{boltztrap}.
In Fig.\,\ref{fig:dos_transport_ni0.4au0.6_csta}(a) the total DOS calculated using the KKR-CPA method for Ni$_{0.4}$Au$_{0.6}$ and the  Ni$_{11}$Au$_{16}$ supercell is compared. Overall, the DOS obtained using these two approaches are in good agreement. 
The CPA approach smears the DOS peak below the Fermi level, which remains sharp and narrow in the supercell calculations, where bands are infinitely sharp and electron scattering effects are absent. 
DOS above $E_F$, where relatively sharp bands are predicted to exist in KKR-CPA, is almost identical to the DOS obtained in the ordered supercell calculations. 

In Fig.\,\ref{fig:dos_transport_ni0.4au0.6_csta}(b) comparison of calculated thermopower using both CRTA and Kubo formalism is shown. 
Apart form $S(T)$ plot for 40\% Ni, we add the experimental data points for a close composition of Ni$_{0.37}$Au$_{0.63}$ \cite{Garmroudi2023}. 

As expected, the supercell + CRTA thermopower is much smaller in absolute value compared to the KKR-CPA + Kubo-Greenwood result, which again is close to the experimental findings.
This confirms that it is the energy-dependent resonant scattering that abruptly disappears above $E_F$, which makes the Ni-Au alloy such an excellent thermoelectric system.

In addition, Supplemental Material, Fig. S6, shows the comparison of our calculated $S(T)$ curves with the simulations of Ref.~\cite{Garmroudi2023} based on $s-d$ model, where thermopower was estimated mostly using $\tau^{-1} \propto N_d$ rule and calculated densities of states (see \cite{Garmroudi2023} for details). As can be seen, the calculations presented here are much closer to the experimental findings.

% ################################################\
\subsection{Ni$_x$Cu$_{1-x}$ versus Ni$_x$Au$_{1-x}$}
% ################################################\

\begin{figure}
    \includegraphics[width=1\linewidth]{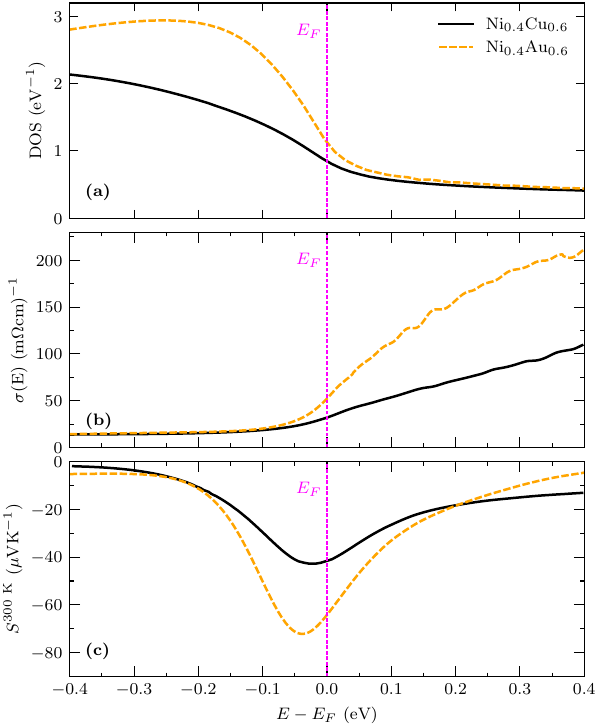}
    \caption{(a) Total density of states, (b) transport function $\sigma(E)$ and (c) calculated energy dependent room temperature  Seebeck coefficient $S^{300\,\mathrm{K}}(E)$ for Ni$_{0.4}$Cu$_{0.6}$ and Ni$_{0.4}$Au$_{0.6}$.}
    \label{fig:niau_vs_cuni_dos_sgm_s}
\end{figure}

\begin{figure}
    \includegraphics[width=1\linewidth]{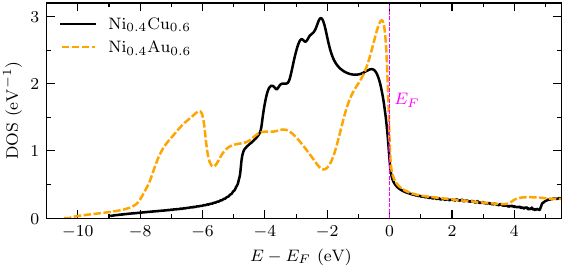}
    \caption{Total density of states for Ni$_{0.4}$Cu$_{0.6}$ and Ni$_{0.4}$Au$_{0.6}$. }
    \label{fig:niau_vs_cuni_dos_total_wide}
\end{figure}

\begin{figure}
    \includegraphics[width=1\linewidth]{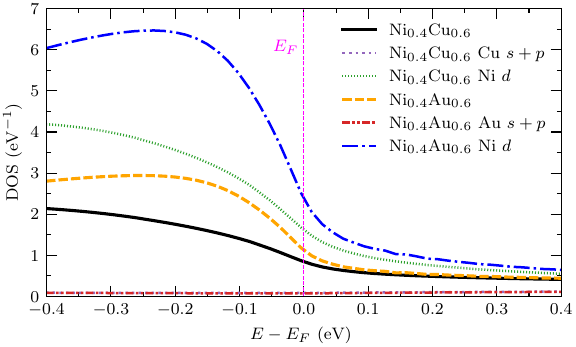}
    \caption{Partial density of states for Ni$_{0.4}$Cu$_{0.6}$ and Ni$_{0.4}$Au$_{0.6}$. }
    \label{fig:niau_vs_cuni_dos_partial}
\end{figure}

\begin{figure*}
    \includegraphics[width=1\linewidth]{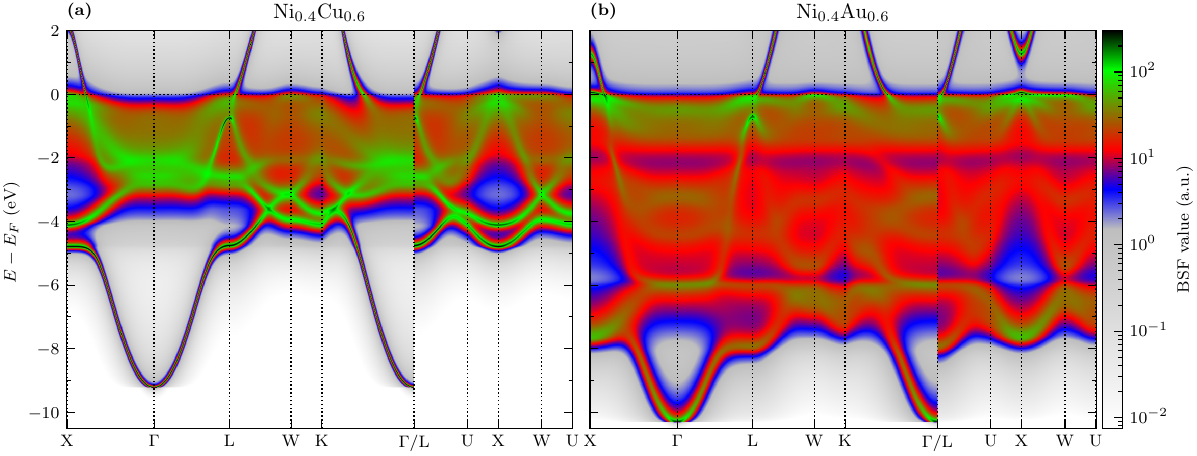}
    \caption{Calculated 2D BSF projections for (a) Ni$_{0.4}$Cu$_{0.6}$   and (b) Ni$_{0.4}$Au$_{0.6}$. BSFs are plotted in logarithmic scale, with values higher than 300 atomic units marked as black color. }
    \label{fig:niau_vs_cuni_bsf}
\end{figure*} 

\begin{figure*}
    \includegraphics[width=1\linewidth]{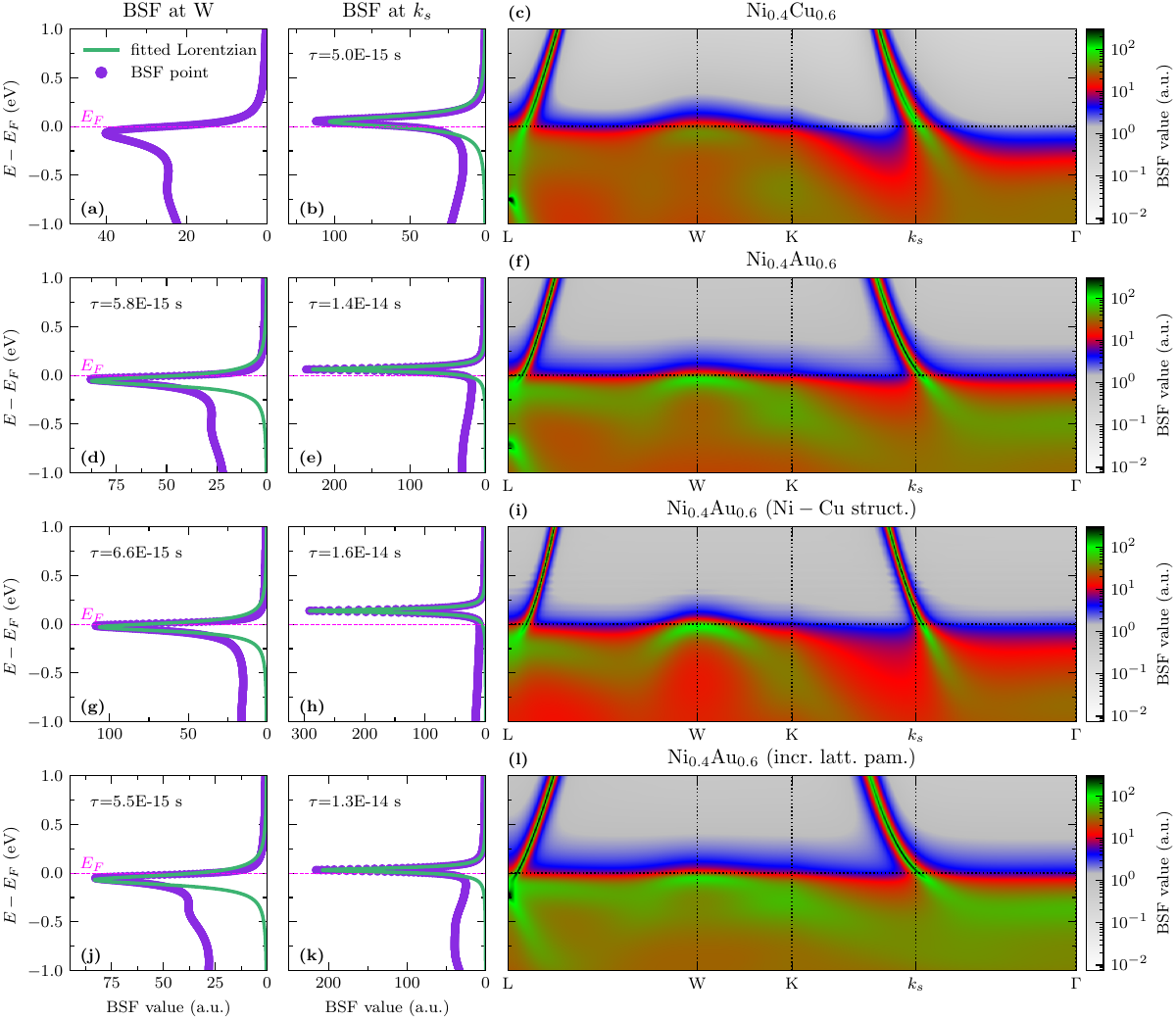}
    \caption{Bloch spectral functions for 
    (a-c) Ni$_{0.4}$Cu$_{0.6}$; 
    (d-f) Ni$_{0.4}$Au$_{0.6}$; 
    (g-h) Ni$_{0.4}$Au$_{0.6}$ calculated in Ni$_{0.4}$Cu$_{0.6}$ structure (reduced lattice parameter to 3.57 \AA); 
    and (j-l) Ni$_{0.4}$Au$_{0.6}$ with increased lattice constant to 4.075\ \AA. 
    Panels (a,d,g,j) and (b,e,h,k) show BSF at a given $\vb{k}$-points: W and $k_s$ ($(0.33, 0.33, 0.00)\frac{2\pi}{a}$) respectively, where $k_s$ is located between K and $\Gamma$ (shown in right panels), and chosen as a point where the band crosses $E_F$. Panels (c,f,i,l) show 2D projections of BSF function, with color representing BSF value in logarithmic scale. Black color corresponds to values higher than 300 atomic units.}
    \label{fig:niau_vs_cuni_bsf_zoom}
\end{figure*}

To understand the unique properties of Ni$_x$Au$_{1-x}$ we compare it here with the reference thermoelectric metallic alloy Ni$_x$Cu$_{1-x}$. Constantan is one of the first metallic alloys in which resonant electronic states on impurity atoms were recognized, and due to its high thermopower it is commonly used in thermocouples.
Its electronic structure has been widely studied experimentally and theoretically \cite{Wiendlocha2018Thermopower,Gordon1981OnAlloys,Vernes2003,Barbieri2004,Bansil1975,Stocks1978,Seib1970a,Seib1970b,Seib1968}, thus here we focus only on the similarities and differences between the two systems.

In Fig.\,\ref{fig:niau_vs_cuni_dos_sgm_s}(a) and Fig.\,\ref{fig:niau_vs_cuni_dos_total_wide} the total DOS of Ni$_{0.4}$Cu$_{0.6}$ and Ni$_{0.4}$Au$_{0.6}$ are presented as well as partial DOS is shown in Fig.\,\ref{fig:niau_vs_cuni_dos_partial}. 
The total DOS curves are much different below the Fermi level, with a larger and steeper peak in the latter. This increase in DOS arises from the much higher 3$d$ Ni states in Ni$_{0.4}$Au$_{0.6}$, as both 3$d$ Cu and 5$d$ Au remain similar.
However, above the Fermi level, the electronic DOSes are almost identical. 
Thus, simplifying that $\sigma(E)\propto 1/N(E)$ one should expect different $\sigma(E)$ functions below $E_F$ and similar above $E_F$.
Counterintuitively, as can bee seen in Fig.\,\ref{fig:niau_vs_cuni_dos_sgm_s}(b), this is not the case, as in the energy range where both alloys possess the RL the $\sigma(E)$ is equal, and only begins to differ above the Fermi level, where the DOS of both alloys is identical, leading to nearly double values of the transport function at 0.4 eV. 
This steeper $\sigma(E)$ is responsible for having a much higher absolute value of $S$ in Ni-Au (Fig.\,\ref{fig:niau_vs_cuni_dos_sgm_s}(c)) with nearly 30 $\mu$V/K more at $E_F$.

This shows that the energy dependence of the conductivity, $\sigma(E)$, is not purely arising from the density of states, but is shaped by the difference in electronic scattering. 
To explain this further, the Bloch spectral density functions of constantan and Ni$_{0.4}$Cu$_{0.6}$ are compared in Fig.\,\ref{fig:niau_vs_cuni_bsf}(a,b). 
In a broad picture, the electronic structure of Ni-Cu is characterized by sharper bands than in the Ni-Au alloy, where the spectral functions are highly smeared. In the energy region close to $E_F$, the bands became quite similar. 
However, if we focus on a narrow energy range, as shown in Fig.\,\ref{fig:niau_vs_cuni_bsf_zoom}(c,f), it becomes clear that the band structure in Ni$_{0.4}$Au$_{0.6}$ at $E_F$ and just above $E_F$ is characterized by sharper and narrower Bloch spectral functions. 

In Fig.\,\ref{fig:niau_vs_cuni_bsf_zoom}(a,b,d,e) BSFs for selected $\vb{k}$-points (W and $k_s\sim(0.33, 0.33, 0.00)\frac{2\pi}{a}$) are plotted as a function of energy. Furthermore, a fit to the Lorentz function Eq.\,\eqref{eq:lorentz} is added as a green line. 
Apart form the case of BSF at the W point for the Ni-Cu alloy, where fitting was not possible due to the large 'shoulder', the lifetimes of the electronic states ($\tau=\hbar/\Delta$) were calculated using the width $\Delta$ of the spectral function from the obtained fits. For Ni-Au at W point (Fig.\,\ref{fig:niau_vs_cuni_bsf_zoom}(d)) it was possible to establish $\Delta=0.12$ eV, which corresponds to $\tau=5.8\cdot 10^{-15}$ s. 
When considering BSF at $k_s$ point, we obtain a considerable difference between both systems from $\tau=5.0\cdot 10^{-15}$~s for Ni-Cu to $\tau=1.4\cdot 10^{-14}$~s for NiAu, a twice longer electronic lifetime. An increase in electronic lifetime indicates weaker scattering in the Ni-Au alloy. If we consider the Drude formula $\sigma=\frac{ne^2\tau}{m^*}$, where $n$ is the carrier concentration and $m^*$ is the effective mass, we can see the cause of Ni-Au superiority, as a longer electronic lifetime directly enhances~$\sigma$. This means that the steep $\sigma(E)$ in Ni-Au and, as a consequence, the large thermopower, is actually caused by a combination of the presence of a flat and strongly smeared (due to resonant scattering) band below $E_F$, similar to that in Ni-Cu, and a sharp, steep and almost linear band with weaker scattering at and above the Fermi level. 
A similar combination of flat and dispersive bands was observed, for example, in metallic CoSi~\cite{cosi} or recently in Ni$_3$In$_{1-x}$Sn$_x$ alloy \cite{Garmroudi2025Topological}, resulting also in excellent thermoelectric properties.

\subsection{Influence of lattice parameter variation}

\begin{figure}
    \includegraphics[width=1\linewidth]{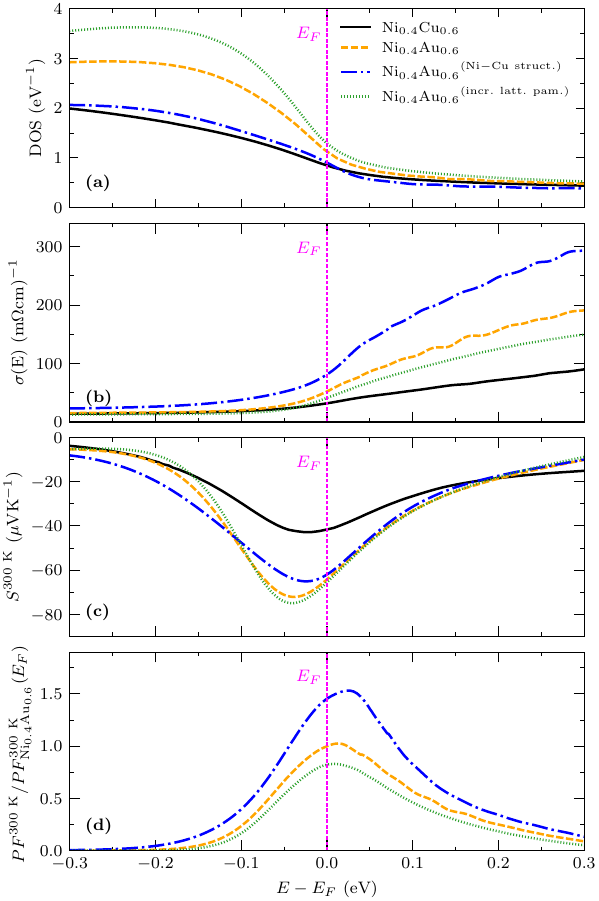}
    \caption{(a) Total density of states; (b) transport function $\sigma(E)$; (c) calculated energy dependent room temperature Seebeck coefficient $S^{300\,\mathrm{K}}(E)$ for Ni$_{0.4}$Cu$_{0.6}$ and Ni$_{0.4}$Au$_{0.6}$ as well as two variants: Ni$_{0.4}$Au$_{0.6}$ calculated in Ni$_{0.4}$Cu$_{0.6}$ structure (reduced lattice parameter to 3.57 \AA) and Ni$_{0.4}$Au$_{0.6}$ with increased lattice constant to 4.075\ \AA. Panel (d) shows calculated ratio of room temperature power factor to $PF^{300\ K}$ of Ni$_{0.4}$Au$_{0.6}$ at the Fermi level. The $PF$ used here were estimated using the calculated $S$ at 300 K and transport function $\sigma(E)$.}
    \label{fig:niau_variants_transport}
\end{figure}

In our quest to understand the differences between Ni-Au and Ni-Cu systems, which may help to further develop more efficient metallic thermoelectric alloys, in addition to the elemental difference in the type of host system (Cu vs. Au) we note that these two materials have a much different lattice parameter. This opened up a question of the role of chemical pressure in shaping the transport properties of Ni-Au. 
To characterize it quantitatively, electronic structure and transport calculations were performed for Ni$_{0.4}$Au$_{0.6}$ system with a modified (decreased and increased) lattice constant. 
For the case of a decreased lattice parameter (positive chemical pressure), it was matched with the lattice parameter of Ni$_{0.4}$Cu$_{0.6}$ (reduction from 3.881~\AA\ to 3.570~\AA).
For the simulation of the negative chemical pressure, the lattice parameter was expanded in 5\%, to 4.075~\AA.

In Fig.\,\ref{fig:niau_vs_cuni_bsf_zoom} and Fig.\,\ref{fig:niau_variants_transport} results of calculations of the electronic structure and transport properties are shown for both cases. 

Starting with the system in the Ni$_{0.4}$Cu$_{0.6}$ structure (Fig.\,\ref{fig:niau_variants_transport} -- blue dash-dotted line), reduction of the lattice parameter has significantly altered the electronic structure, as the DOS below $E_F$ decreased and now appears very similar to Ni$_{0.4}$Cu$_{0.6}$, with just a slightly steeper slope at the Fermi level (Fig.\,\ref{fig:niau_variants_transport}(a)).
Nevertheless, the transport function (Fig.\,\ref{fig:niau_variants_transport}(b)) is much different from the constantan case, which again confirms that it is not the sole density of states that shapes the thermoelectric properties of Ni-Au. 
From $-0.3$ to $0.3$ eV, $\sigma(E)$ of the compressed alloy increased by $50-60$\% in comparison to initial Ni$_{0.4}$Au$_{0.6}$ case. 
The increase in $\sigma(E)$ at $E_F$ appears also to be faster, however, the potential gain in thermopower arising from the rapidly changing transport function was nullified by the increase in $\sigma(E)$ below the Fermi level, leading to an overall slightly lower $S$ (Fig.\,\ref{fig:niau_variants_transport}(c)).

Expanding the lattice parameter of Ni$_{0.4}$Au$_{0.6}$ (Fig.\,\ref{fig:niau_variants_transport} -- green dotted line) has an entirely opposite effect, with the DOS peak being visibly higher. 
The shape of the transport function remains mostly unaffected except for its values being $\sim20$ \% lower compared to 'real' Ni$_{0.4}$Au$_{0.6}$. 
In terms of thermopower, only a small increase of 2\% is present at $E_F$.

Observed behavior, i.e. the general increase or decrease of $\sigma(E)$ in the two above-mentioned examples, is directly related to the sharpening or smearing of spectral functions, especially below the Fermi level. This is indicated in both the 2D projections of the Bloch spectral functions in compressed (Fig.\,\ref{fig:niau_vs_cuni_bsf_zoom}(i)) and expanded (Fig.\,\ref{fig:niau_vs_cuni_bsf_zoom}(l)) of  Ni$_{0.4}$Au$_{0.6}$,t as well as the BSF profiles which, e.g., for the expanded structure show much larger 'shoulders' below the sharp peak at $E_F$.

Fig.\,\ref{fig:niau_variants_transport} (d) shows the room-temperature power factor normalized to the value of $PF^{\mathrm{300\ K}}$ of Ni$_{0.4}$Au$_{0.6}$ at the Fermi level. 
Here, the $PF$ was estimated using the Seebeck coefficient calculated via the Fermi integrals (Eqs.\,(\ref{eq:seebeck},\ref{eq:ln})) and the energy-dependent transport function $\sigma(E)$. Because the obtained $\sigma(E)$ does not include the effects of the electron-phonon scattering (making the reliable calculation of absolute values of conductivity impossible), we show only a relative effect which demonstrates that by local expansion of contraction of the lattice parameters, e.g. by alloying with smaller or larger elements or by deposition of thin layers on a mismatched substrate, it should be possible to influence the thermopower and power factor. 
Certainly such large variations of the lattice parameter as used here in the simulation are not possible with any small addition of elements, but this route towards optimization of the thermoelectric performance of alloys should also be considered. 

From the two variants studied, a potential enhancement of the power factor could be achieved by decreasing the lattice parameter.
Despite the initial reduction of $S$, the increase of $\sigma(E)$ for Ni$_{0.4}$Au$_{0.6}$ in Ni$_{0.4}$Cu$_{0.6}$ structure compensates for it, increasing $PF$ almost 50\% close to the Fermi level.
An increase in thermopower can be expected for an expanded structure. 

\section{Summary \& Conclusions}
In summary, we performed a theoretical study of the electronic structure and transport properties of Ni-Au metallic alloys using the KKR-CPA method. Our results show that Ni doping induces a resonant level below the Fermi level, similar to what is observed in Ni-Cu alloys.

As the Ni concentration increases, this resonant state evolves into a virtual, flat, and strongly smeared band below the Fermi level. This band hybridizes with the steep linear band originally present in Au, leading to its breaking below $E_F$. Consequently, the energy-dependent conductivity $\sigma(E)$ decreases below $E_F$. Simultaneously, due to the energy-selective nature of resonant scattering, which is absent above $E_F$, the linear band remains largely intact above $E_F$ and experiences only weak broadening.

This asymmetric scattering results in a steep increase in $\sigma(E)$ as the energy crosses $E_F$, producing a large thermopower in Ni-Au alloys near equiatomic composition.

The highest room-temperature thermopower of approximately $-75$ $\mu$V/K is achieved at 50\% Ni, while alloys with 60\% Ni would reach nearly $-101$ $\mu$V/K at 700~K. The prime role of electron scattering is further supported by comparison with calculations based on periodic supercell + CRTA calculations: Despite similar densities of states, neglecting energy-dependent electron scattering yields significantly lower Seebeck coefficients than those obtained when scattering is taken into account within the KKR-CPA method.

A comparative study with constantan (Ni-Cu alloy) confirms that the superior thermoelectric properties of Ni$_x$Au$_{1-x}$ are not merely due to the steep density of states near $E_F$, which is similar in both systems. Instead, the key difference lies in the energy dependence of the conductivity $\sigma(E)$, driven by differences in electron scattering. Spectral function analysis shows that Ni$_x$Au$_{1-x}$ exhibits roughly twice the electronic lifetime near $E_F$ compared to Ni-Cu. This results in a sharper transition from strong resonant scattering below $E_F$ to weak scattering above, which underpins the high thermoelectric performance of the Ni-Au system.

Finally, our analysis of lattice parameter variations for $x=0.4$ indicates that tuning the lattice constant can effectively influence the thermopower and power factor. This suggests a promising strategy for further enhancement, such as alloying with elements that reduce the lattice constant or depositing thin films on substrates with smaller lattice parameters.

% ################################################\
\begin{acknowledgments}

Research project supported by the 'Excellence Initiative – Research University' program at AGH University of Krakow.
The authors gratefully acknowledge the Polish high-performance computing infrastructure PLGrid (HPC Center: ACK Cyfronet AGH) for providing computer facilities and support within computational grant no. PLG/2025/018404.

\end{acknowledgments}

% \appendix

% \section{Appendixes}

\newpage

\bibliography{ref}

\vspace*{40pt}

\newpage
\onecolumngrid
\section*{Supplemental Material}

\renewcommand{\thefigure}{{S\arabic{figure}}}
\setcounter{figure} 0

\begin{figure*}[h]
    \includegraphics[width=0.95\linewidth]{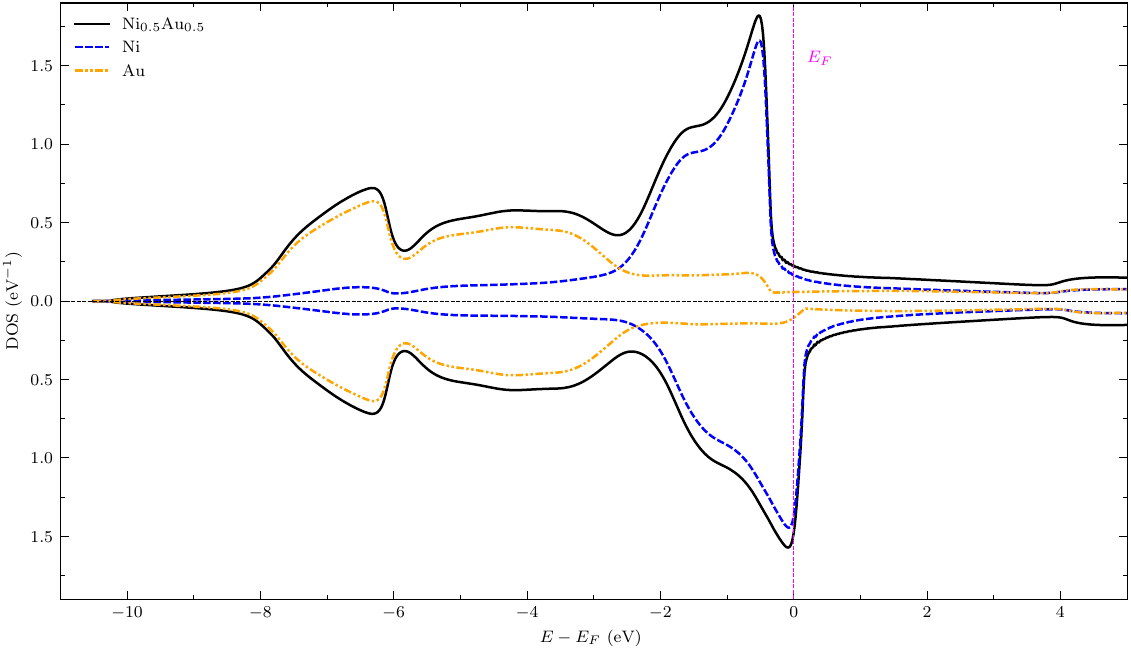}
    \caption{Calculated spin-polarized density of states for Ni$_{0.5}$Au$_{0.5}$ with atomic contribution. }
    \label{fig:sup:dos_ni0.5au0.5_sprkkr_mag}
\end{figure*}

Figure\,\ref{fig:sup:dos_ni0.5au0.5_sprkkr_mag} shows spin-polarized density of states for magnetic calculation of Ni$_{0.5}$Au$_{0.5}$. Here, the computational details remain unchanged from those discussed in the paper apart from an increase in the number of $\vb{k}$-points to $1.6\cdot10^4$ in the self-consistent-field cycle. The overall shape of DOS in this system remains similar to Ni$_{0.5}$Au$_{0.5}$ in non-magnetic calculation, with main difference being localization of the Fermi level, which in majority spin channel lies now at the bottom of the DOS peak. This results in a total magnetic moment per unit cell of 0.226 $\mu_B$.

% In Fig.\,\ref{fig:sup:transport_sigma},  Fig.\,\ref{fig:sup:transport_sigma_zoom} and Fig.\,\ref{fig:sup:transport_seebeck} results of calculated energy-dependent transport function $\sigma(E)$ and temperature-dependent Seebeck coefficient are presented for all studied concentrations of Ni$_{1-x}$Au$_x$.

\begin{figure*}[h]
    \includegraphics[width=0.95\linewidth]{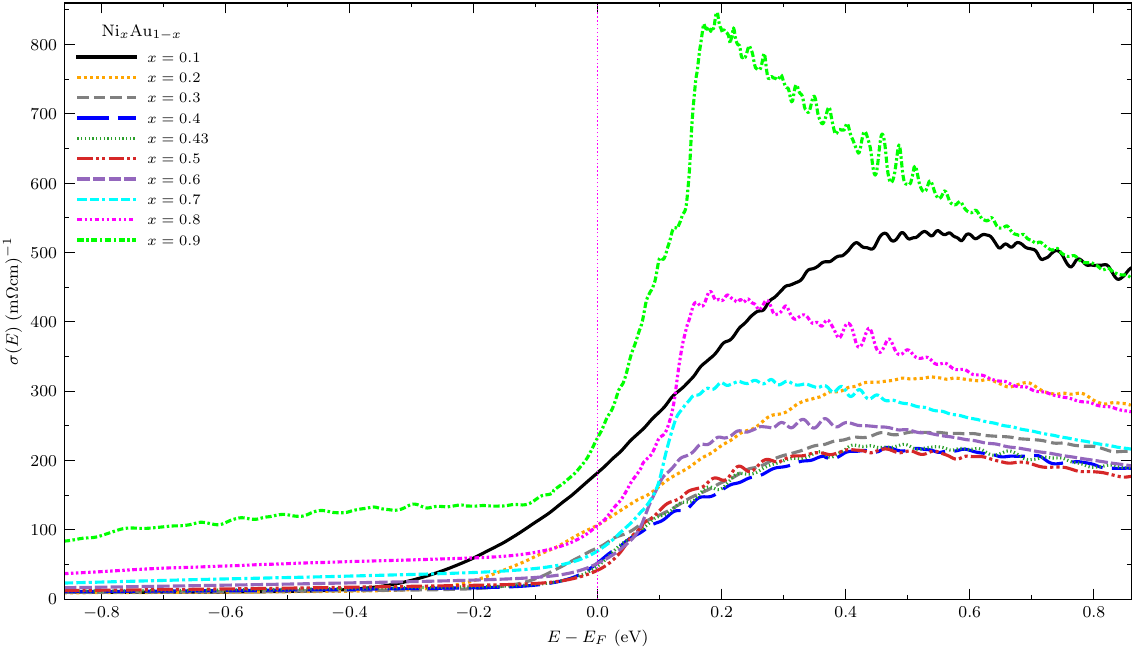}
    \caption{Energy-dependent transport function $\sigma(E)$ of Ni$_x$Au$_{1-x}$ for $x=0.1-0.9$.}
    \label{fig:sup:transport_sigma}
\end{figure*}

\begin{figure*}[h]
    \includegraphics[width=0.95\linewidth]{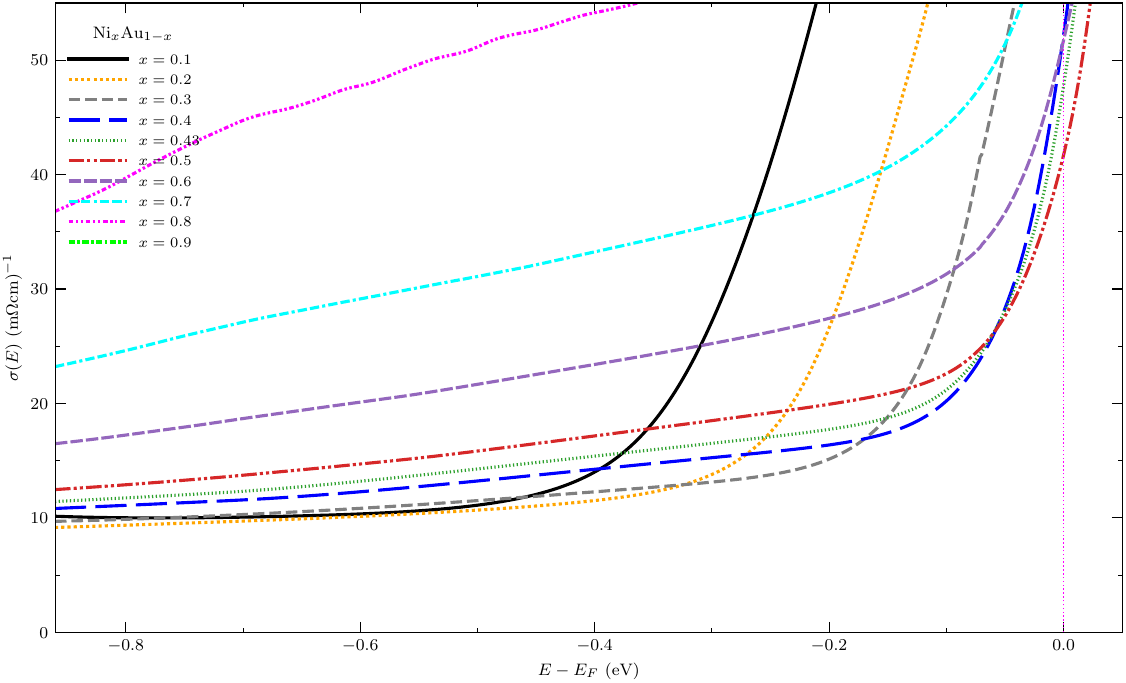}
    \caption{Energy-dependent transport function $\sigma(E)$ of Ni$_x$Au$_{1-x}$ for $x=0.1-0.9$ in the energy region below the Fermi level. }
    \label{fig:sup:transport_sigma_zoom}
\end{figure*}

\begin{figure*}[h]
    \includegraphics[width=0.95\linewidth]{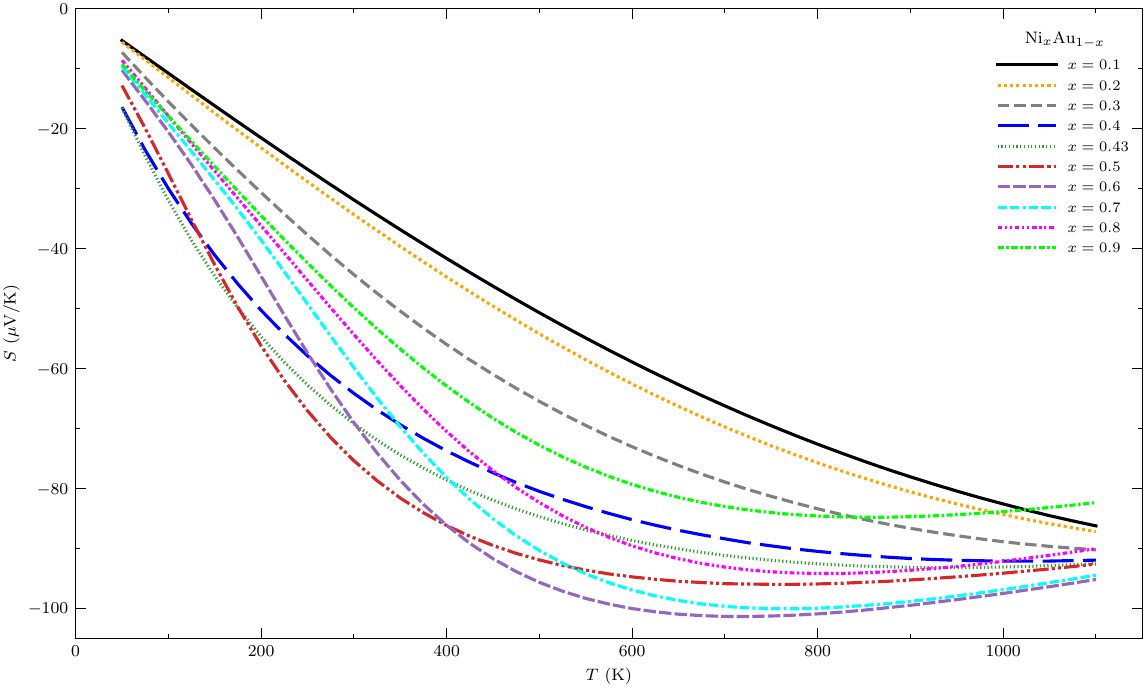}
    \caption{Calculated Seebeck coefficient as a function of temperature $S(T)$ of Ni$_x$Au$_{1-x}$ for $x=0.1-0.9$.}
    \label{fig:sup:transport_seebeck}
\end{figure*}

In Fig.\,\ref{fig:sup:transport_sigma} and Fig.\,\ref{fig:sup:transport_sigma_zoom} results of calculated transport function $\sigma(E)$ are presented for all studied concentrations of Ni$_{1-x}$Au$_x$. The transport function in the presented energy range consists qualitatively of two different regions. The first one, with relatively small conductivity below $E_F$, which corresponds to the energy range with a flat virtual band that is resonantly smeared with increased Ni concentration (region 2 in the BSF discussion). Above the Fermi level, conductivity values rapidly increase as a result of the existence of a sharp, linear band with much weaker smearing (region 3 in the BSF discussion). In Fig.\,\ref{fig:sup:transport_seebeck}, calculated temperature-dependent Seebeck coefficient is shown for all concentrations studied.

\begin{figure*}[h]
    \includegraphics[width=0.95\linewidth]{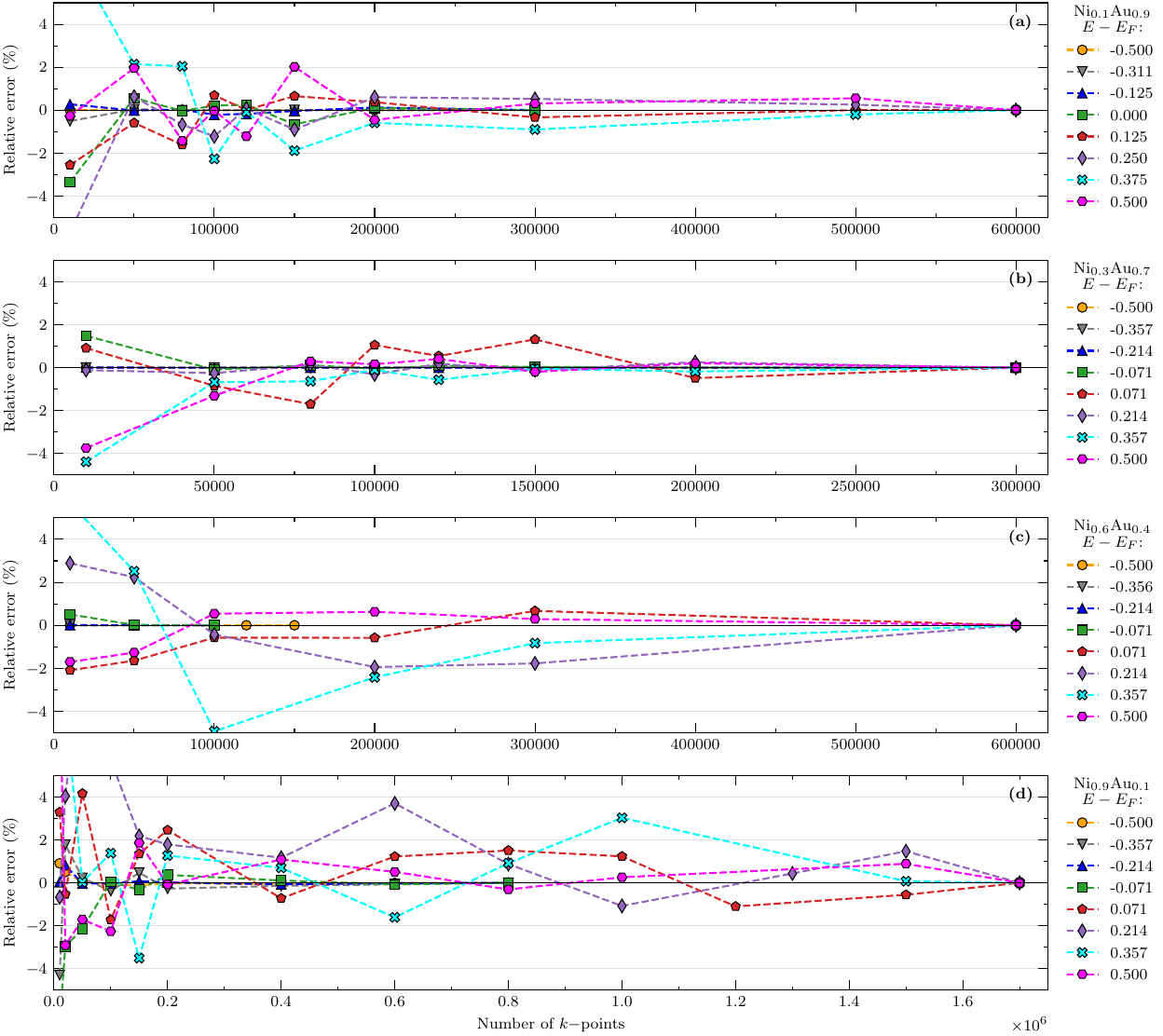}
    \caption{Convergence tests of the transport function $\sigma(E)$ for (a) Ni$_{0.1}$Au$_{0.9}$, (b) Ni$_{0.3}$Au$_{0.7}$, (c) Ni$_{0.6}$Au$_{0.4}$, (d) Ni$_{0.9}$Au$_{0.1}$. Markers correspond to points calculated at different energy with respect to the Fermi level.}
    \label{fig:sup:sigma_convergence}
\end{figure*}

Fig.\,\ref{fig:sup:sigma_convergence}(a-d) presents results of $\sigma(E)$ convergence tests for $x=0.1,0.3,0.6,0.9$. The transport function was calculated at given energy for a different number of $\vb{k}$-points and here is shown the relative error with respect to the results obtained for meshes with the highest number of $\vb{k}$-points. Markers represent tests performed for different energies in each system with reference to the Fermi level $E_F$.

\begin{figure*}[h]
    \includegraphics[width=0.95\linewidth]{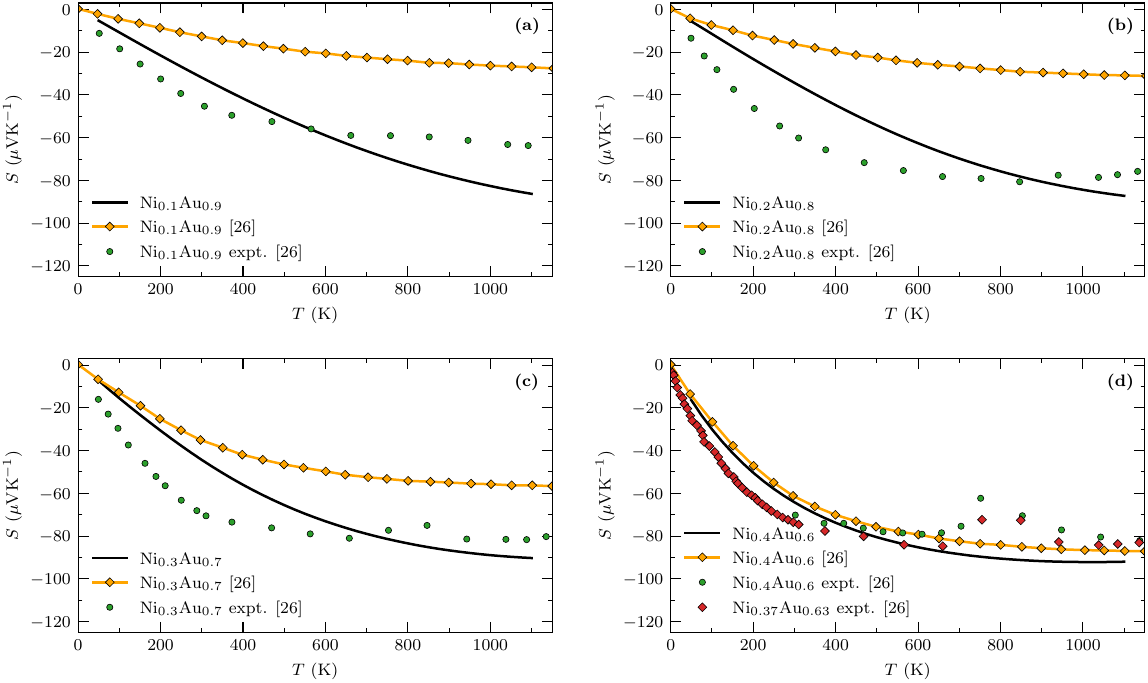}
    \caption{Calculated temperature-dependent thermopower of $\mathrm{Ni_xAu_{1-x}}$ for $x=0.1,0.2,0.3,0.4$. In addition, results from [26] are shown: experimental $S(T)$ as well as calculated thermopower. For $x=0.3, 0.4$ experimental data are shown for measurement obtained only during heating process due to phase segregation occurring at high temperatures. Phase segregation was also responsible for the non-monotonic behavior of measured thermopower above 600 K in $x = 0.37$ and $0.4$ sample [26] in panel (d).}
    \label{fig:sup:thermopower_models_comp}
\end{figure*}

In Fig.\,\ref{fig:sup:thermopower_models_comp}(a-d), comparison of our calculated temperature-dependent Seebeck coefficient with the results of the modeling of [26] is shown. Experimental results are also included. 
For $x \geq 0.30$ experimental data are shown for measurements obtained only during heating, as phase segregation occurs at high temperatures. 
Phase segregation is also responsible for the non-monotonic behavior of measured thermopower above 600 K in $x = 0.37$ and $0.4$ samples [26] in panel (d).
The theoretical thermopower in [26] was obtained based on the $s-d$ model, where the thermopower was estimated using the $\tau^{-1} \propto N_d$ rule and the calculated densities of states (see [26] for details).

Table\,\ref{tab:sup:lambda} contains experimental electronic specific heat coefficients $\gamma_{\rm expt}$ from [64], theoretical values $\gamma_{\rm band}$ calculated using the KKR-CPA density of states $N(E_F)$ and renormalization parameters $\lambda = \gamma_{\rm expt}/\gamma_{\rm band} -1$, which in the absence of other interactions is the electron-phonon coupling parameter. 

\vfill

\begin{table}[p]
\caption{Experimental electronic specific heat coefficients of Ni$_{x}$Au$_{1-x}$ $\gamma_{\rm expt}$ from [64], theoretical values $\gamma_{\rm band}$ calculated using the KKR-CPA density of states $N(E_F)$ and renormalization parameters $\lambda = \gamma_{\rm expt}/\gamma_{\rm band} -1$, which in the absence of other interactions is the electron-phonon coupling parameter.}
\label{tab:sup:lambda}
\begin{ruledtabular}
\begin{tabular}{ccccc}
$x$   & $\gamma_{\rm expt.}$ & DOS (eV$^{-1}$)   & $\gamma_{\rm band}$ (mJ mol$^{-1}$ K$^{-2}$) & $\lambda$ \\ \hline
0.2 & 1.5         & 0.52      & 1.226 & 0.223  \\
0.3 & 1.8         & 0.68       & 1.603 & 0.123  \\
0.4 & 4.0           & 1.16      & 2.735 & 0.462 
\end{tabular}
\end{ruledtabular}
\end{table}

\end{document}